\documentclass{IEEEtran}
\IEEEoverridecommandlockouts
\usepackage{cite}
\usepackage{amsmath,amssymb,amsfonts}
\usepackage{algorithmic}
\usepackage{graphicx}
\usepackage{textcomp}
\usepackage{xcolor}

\newcommand{\fref}[1]{Fig.~\ref{#1}}

\def\BibTeX{{\rm B\kern-.05em{\sc i\kern-.025em b}\kern-.08em
    T\kern-.1667em\lower.7ex\hbox{E}\kern-.125emX}}
\begin{document}

\title{Upstream Allocation of Bidirectional Load Demand\\by Power Packetization}

\author{Shiu~Mochiyama, Kento~Hiwatashi, and~Takashi~Hikihara%
\thanks{This work was partially supported by JSPS KAKENHI 24K17262, Kayamori Foundation of Informational Science Advancement, and Kansai Research Foundation for Technology Promotion.}%
\thanks{The authors are with Kyoto University, Kyoto 615-8510 Japan. Correspondence to: s-mochiyama@dove.kuee.kyoto-u.ac.jp}}

\maketitle

\begin{abstract}
The power packet dispatching system has been studied for power management with strict tie to an accompanying information system through power packetization. 
In the system, integrated units of transfer of power and information, called power packets, are delivered through a network of apparatuses called power packet routers. 
This paper proposes upstream allocation of a bidirectional load demand represented by a sequence of power packets to power sources.
We first develop a scheme of power packet routing for upstream allocation of load demand with full integration of power and information transfer. 
The routing scheme is then proved to enable packetized management of bidirectional load demand, which is of practical importance for applicability to, e.g., electric drives in motoring and regenerating operations. 
We present a way of packetizing the bidirectional load demand and realizing the power and information flow under the upstream allocation scheme. 
The viability of the proposed methods is demonstrated through experiments.
\end{abstract}


\section{Introduction}

\begin{figure*}[tb]
\centering
\includegraphics[width=17cm]{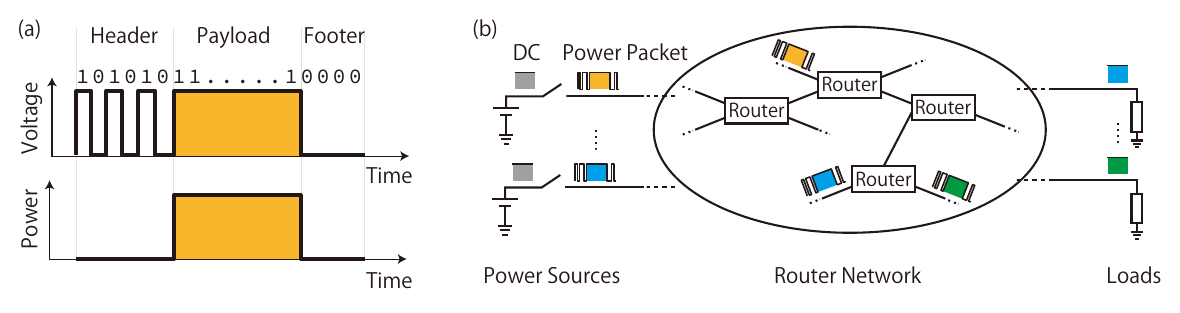}
\caption{Power packet dispatching system \cite{Takahashi.etal-2015,Mochiyama.etal-2021}. (a) Example configuration of a power packet. The \texttt{1} and \texttt{0} above the voltage waveform represent high and low logic states, respectively. 
(b) Example of a power packet dispatching system with a network of power packet routers. }
\label{fig:ppds}
\end{figure*}

As a means of energy management in a system disconnected from a large-scale power source (hereafter, we denote an isolated system), there is a proposal of the power packet dispatching system \cite{Takahashi.etal-2015,Mochiyama.etal-2021,Mamiya.etal-2024}. 
In the system, the power is packetized by dividing the energy flow into power pulses with information tags attached physically (Fig.~\ref{fig:ppds}~(a)). 
Packetized power is routed from a source to a load through a network of apparatuses called power packet routers according to the tag (Fig.~\ref{fig:ppds}~(b)). 
The authors' group has developed a physical realization of a power-packet router and verified its operation in networked routers through experiments\cite{Takahashi.etal-2015,Katayama.Hikihara-2020a,Yoshida.etal-2020,Mochiyama.etal-2021}. 

The key enablers of the concept are the division of power in the time domain and the attachment of the physical tag that indicates the origin and destination of each power pulse. 
The simultaneity between physical and cyber quantities is the most essential aspect in such a cyber-physical system \cite{Kim.Kumar-2012,Guo.etal-2017,Zhao.etal-2024}. 
For example, in a typical situation in isolated systems with increased harvesting sources, deployment of batteries of enormous capacity is not realistic due to the limited cost, space, and weight. 
It is not easy to balance instantaneous supply and demand, both of which have unpredictable and varying profiles, without a large buffer. 
Physical packetization gives complete traceability to power transfer, enabling the routing network to allocate supply and demand between sources and loads with arbitrary proportion counted by the packetized units.
This realizes best-effort matching of source capacity and load demand in a decentralized way, like the Internet. 

From the viewpoint of the loads, the power packet dispatching system delivers power as a collection of discrete units.
Thus, the load control at the edge of the network employs a discretized power processing method. 
This is quite different from the standard way in power electronics, namely pulse width modulation \cite{holmes2003pulse}.
Previous studies\cite{Takahashi.etal-2016a,Mochiyama.Hikihara-2019a} proposed a method to derive an optimal packet sequence to satisfy a load demand. 
The derived sequence is then shared with the routing network as a demand signal, and for each packet in the sequence, a source that meets the requirement generates a power packet to supply the load. 
In this sense, the method performs an upstream allocation of the discretized load demand to sources in the routing network \cite{Bialek-1996,Nawata.etal-2016}. 

However, a remaining issue of the previous studies was that the transfer of power and information was not completely closed in the physical layer. 
That is, the allocation of load demand was calculated at the edge of the routing network and then delivered to the source-side router not in packetized form but via another channel. 
Obviously, this brings about a flaw in the concept of integration of power and information transfer. 

This paper discusses an upstream allocation of load demand that fully integrates the transfer of both power and information as a power packet. 
The contribution of this paper is threefold. 
First, we propose a routing method of power packets (Section~\ref{sect:routing}) where the load demand information is allocated by an information tag to the demanded power source in the upstream direction. 
The pivotal factor of the proposal is the elimination of an implicit assumption of the previous studies that the direction of power and information delivery coincides for a particular power packet. 
The proposed method allows a request from a load (information) and a response from a source (power) to coexist as a single power packet.
Second, we present that the proposed routing method also offers a practical advantage in control of loads with bidirectional power flow, such as powering and regenerating operations of electric drives. 
The previous methods\cite{Takahashi.etal-2016a,Mochiyama.Hikihara-2019a} only apply to the allocation of the unidirectional load demand. 
We propose a method for the upstream allocation of bidirectional load demand (Section~\ref{sect:src_select}), which realizes a seamless handling of a bidirectional power packet flow with completely integrated power and information transfer. 
Lastly, we demonstrate the viability of the proposed method in terms of the two aspects mentioned above through experiments (Section~\ref{sect:exper}). 

\section{Proposed Method}
\label{sect:method}

\subsection{Description of Target System}
\begin{figure*}[tb]
\centering
\includegraphics[width=17cm]{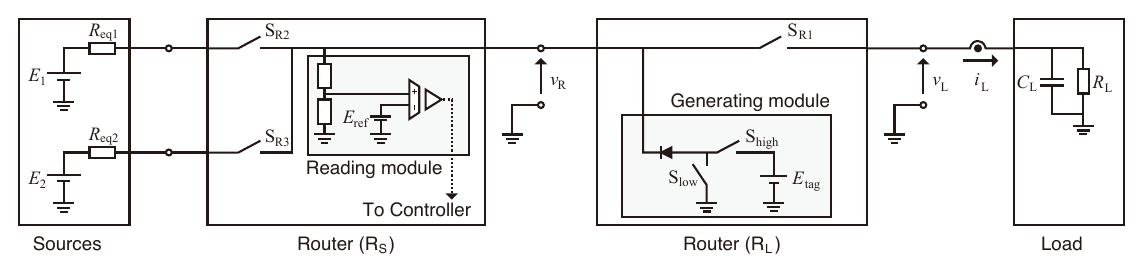}
\caption{Circuit configuration of the power packet dispatching system. }
\label{fig:circ}
\end{figure*}

Fig.~\ref{fig:circ} depicts the power packet dispatching system we consider in this paper. 
We focus on a part of a whole routing network that includes two power sources that supply a particular load. 
Note that the power sources in this paper include ones with sink operation, such as batteries, to consider the bidirectional management of the packetized power supply. 
In fact, we assume that one source is a pure source and the other a rechargeable battery in the experimental setups in Section~\ref{sect:exper}. 
The two-source setup is minimal but generic for discussing bidirectional and upstream allocation. 
The generality and scalability of this configuration will be discussed later in Section~\ref{sect:discuss}.

We introduce a power packet configuration protocol that is shared by the whole network. 
Fig.~\ref{fig:ppds}~(a) presents the bit assignment of a power packet. 
In this paper, we fix the time duration of one bit and the bit length of the information tag and payload. 
We set a fixed sequence for the three bits from the beginning of the information tag, namely \texttt{101}, to declare the start of a power packet. 
The remaining three-bit signal expresses the index of the demanded source. 
Of course, the bit length and their assignment are arbitrary and can be modified according to the system requirements. 

The transfer of a power packet is assumed to be synchronized throughout the routing network. 
Based on the assumption, we define an index of time slot $k$ ($k=1,2,3,\dots$) as a representation of the (continuous) time interval
\begin{equation}
    t \in [(k-1)T_{\rm packet}, kT_{\rm packet}),
\end{equation}
where $T_{\rm packet}$ coincides with the time duration of a power packet. 
Thus, one power packet is transferred in one time slot. 

\subsection{Routing Method for Upstream Allocation}
\label{sect:routing}

With the above setups, we develop a method to allocate load demand by upstream dispatching of power packets. 
Here, we focus on the way of power and information transfer as a power packet and ignore how the source of origin is selected for each power packet.
The source selection algorithm will be discussed in Section~\ref{sect:src_select}. 

\begin{figure}[tb]
\centering
\includegraphics[width=9cm]{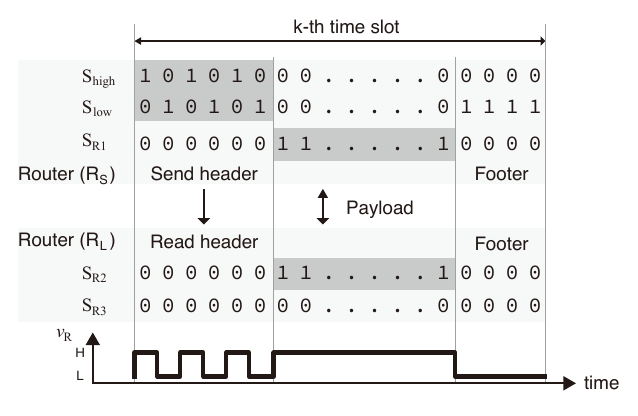}
\caption{Procedure of routers' operation at $k$-th time slot. It is assumed that the source 1 is selected for the depicted time slot. }
\label{fig:procedure}
\end{figure}

For the $k$-th time slot, the router connected to the load and the routers connected to the power sources exchange power and information in a packetized form. 
Fig.~\ref{fig:procedure} presents the overview of router operation throughout a power packet transfer. 
The details will be explained step by step below. 

First, the header tag is sent from the load-side router to the source-side router. 
The sender-receiver relationship is fixed regardless of the direction of the power supply in the bidirectional load case. 
The direction of tag transfer is determined by hardware setups;
the load-side router is equipped with a signal generation module, and the source-side routers are equipped with a signal reading module. 
The signal generation module is made up of a voltage source dedicated to signal generation and two switches controlled in a complementary manner\cite{Katayama.Hikihara-2020a}. 
That is, $({\rm S_{high}},{\rm S_{low}})=({\rm ON}, {\rm OFF})$ generates a voltage for high logic and $({\rm S_{high}},{\rm S_{low}})=({\rm OFF}, {\rm ON})$ pulls down the potential to zero for low logic. 
The source-side routers observe the potential on the transmission line at every bit through the signal reading module comprising a potential divider and a galvanic-isolation comparator\cite{Takahashi.etal-2015}. 
The module converts the potential signal to a logic sequence and passes it to the router controller. 

Second, power transfer is controlled according to the tag information. 
After the load-side router finishes sending the header, it turns on the switch for power transfer, ${\rm S_{R1}}$. 
At the same time, one of the switches of the source-side router, ${\rm S_{R2}}$ or ${\rm S_{R3}}$, specified in the tag, also turns on its switch for power transfer. 
The conduction path between the specified power source and the load is formed in this way. 

Third, the footer follows the payload transfer to indicate the end of one power packet. 
After a predetermined duration of payload passes, the switches of both the load-side and source-side routers are turned off. 

\subsection{Algorithm for Packetization of Bidirectional Load Demand}
\label{sect:src_select}

\begin{figure}[tb]
\centering
\includegraphics[width=8cm]{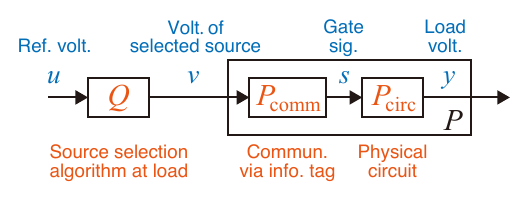}
\caption{Overview of the proposed load control method. }
\label{fig:control}
\end{figure}

In this subsection, we explain how we determine the power source of origin for each time slot. 
We apply a technique for signal quantization\cite{Azuma.Sugie-2008a} to our discrete power processing, based on previous studies\cite{Takahashi.etal-2016a,Mochiyama.Hikihara-2019a}. 
Fig.~\ref{fig:control} shows the overview of the load control we consider. 
Below, we give descriptions for each signal and block depicted in the figure. 

The plant $P$ is composed of two elements: $P_\mathrm{circ}$, the electric circuit depicted in Fig.~\ref{fig:circ}, and $P_\mathrm{comm}$, the communication between the routers through the information tag. 
The circuit includes transistors, and the input of $P_\mathrm{circ}$ is a vector of their switching states. 
Since the load control algorithm only concerns the selection of the source, the input is defined as $\boldsymbol{s}(k) := [\mathrm{S_{R2}},\mathrm{S_{R3}}]^{\mathsf{T}}$.
Then, for the circuit depicted in Fig.~\ref{fig:circ}, applying an appropriate discretization to the circuit equation with a time interval equal to the packet duration $T_{\rm packet}$ yields the following discrete-time system
\begin{equation}
    P_\mathrm{circ}:
    \begin{cases}
        x(k+1) = A x(k) + B V^{\mathsf{T}} \boldsymbol{s}(k) \\
        y(k) = C x(k)
    \end{cases},
\end{equation}
where $x\in\mathbb{R}$ and $y\in\mathbb{R}$ are the state and output, respectively, $A\in\mathbb{R}$, $B\in\mathbb{R}$, and $C\in\mathbb{R}$ are the constant coefficients including the circuit parameters, and $V$ is a vector of possible voltage levels, namely 
\begin{equation}
    \boldsymbol{V}:=[E_1,E_2]^{\mathsf{T}}.
    \label{eq:vlevel}
\end{equation} 
For the setup of Fig.~\ref{fig:circ}, the state and the output are both the voltage of the load, and thus $C=1$.
The product $\boldsymbol{V}^{\mathsf{T}} \boldsymbol{s}(k)$ represents the voltage of the source that is connected to the load at the time interval $k$. 
The switching state $\boldsymbol{s}(k)$ is determined at the source-side router by reading the information tag. 
This operation maps the selection of the voltage level $v\in\{E_1,E_2\}$ to $\boldsymbol{s}(k)$, namely
\begin{equation}
    P_\mathrm{comm}:
    \boldsymbol{s}(k) = 
    \begin{cases}
        [1,0]^{\mathsf{T}} \ \text{if} \ v=E_1 \\
        [0,1]^{\mathsf{T}} \ \text{if} \ v=E_2
    \end{cases}.
\end{equation}
The series conncetion $P_\mathrm{comm}$ and $P_\mathrm{circ}$ is called $P$, the plant in the broad sense. 

The selection of the power source in each time slot is determined by the quantizer $Q$, the algorithm implemented in the load-side router's controller. 
To understand the operation of $Q$, we first consider a reference system $P'$, which is a replacement of $V\boldsymbol{s}(k)$ by a continuous-valued input $u(k)\in\mathbb{R}$ in the original plant $P$. 
$u(k)$ can be recognized as an ideal input, and the quantizer $Q$ seeks to approximate it with a discretized input in the form of power packets. 
The quantizer generates a sequence $\{v(k)\}:=\{v(0),v(1),v(2),\dots v(k)\}$ that makes the output sequence $\{y(k)\}$ as close as possible to the output sequence of the reference system $P'$ under an input sequence $\{u(k)\}$. 
To provide such a quantizer, we consider a dynamic quantizer in the following form
\begin{equation}
    Q:
    \begin{cases}
        \xi(k+1) = A_Q \xi(k) + B_Q (v(k)-u(k)) \\
        v(k) = \boldsymbol{V} \boldsymbol{s}(k) = q(C_Q \xi(k) + u(k))
    \end{cases},
    \label{eq:quantizer}
\end{equation}
where $\xi\in\mathbb{R}$ is the state of the quantizer, and $A_Q\in\mathbb{R}$, $B_Q\in\mathbb{R}$, and $C_Q\in\mathbb{R}$ are the design parameters. 
The function $q(\cdot)$ is a static quantizer that maps its argument to a set of available voltage levels of the system. 
As recognized in (\ref{eq:quantizer}), the quantizer's state stores the weighted sum of past errors between the ideal input $u(k)$ and the actual input $v(k)$.
The output equation of $Q$ then takes this error into account in the argument of the static quantizer at the second equation of (\ref{eq:quantizer}). 
In this way, $Q$ generates $\{v(k)\}$ that dynamically compensates for the quantization error. 
The previous study\cite{Azuma.Sugie-2008a} provides the analytical design of $Q$ for optimal compensation:
\begin{align}
    A_Q &= A, \\
    B_Q &= B, \\
    C_Q &= -A/B.
\end{align}

The bidirectionality of power delivery is a result of the selection of possible voltage levels of the power sources. 
Specifically, setting both higher and lower voltage levels than the nominal value of $u(k)$ leads to a bidirectional current flow. 
For example, when $u(k)$ is in a decreasing trend, the lower voltage source is selected more to draw stored energy from the load, and when it is in an increasing trend, the higher voltage source is selected more to supply energy to the load. 

Lastly, we explain the connection between the proposed routing method and the load demand packetization. 
The output of $Q$ can be assigned to one of the switching states of the plant circuit. 
At the beginning of each power-packet generation, the load-side router calculates $v(k)$. 
The router then assigns the index corresponding to the selected power source to the header. 

\section{Experimental Verification}
\label{sect:exper}

\subsection{Setups}

\begin{figure}[tb]
\centering
\includegraphics[width=9cm]{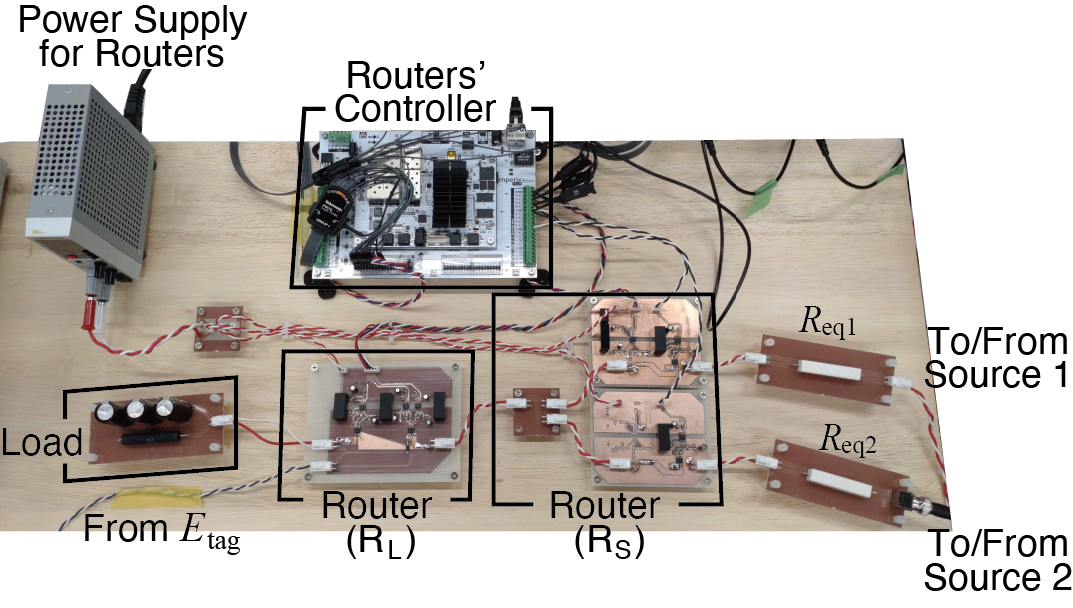}
\caption{Overview of the experimental setups. }
\label{fig:photo}
\end{figure}

We built the routing network depicted in Fig.~\ref{fig:circ} for experiments.
Fig.~\ref{fig:photo} presents an overview of circuit setups. 
The algorithms presented in Section~\ref{sect:method} is implemented in the controller board. 
Although the algorithms for the two routers are implemented into the same controller, they are strictly separated as programs inside the chip and no communication other than clock synchronization is made between the two programs. 
As described in Section~\ref{sect:method}, the routers' communication relies on an information tag through the hardware, tag generating module and tag reading module. 

The voltage values of the sources are set at $E_1 = 12\,\mathrm{V}$ and $E_2 = 3.6\,\mathrm{V}$. 
We assume that source 1 is a pure power source and that source 2 is a battery that accepts a bidirectional current flow. 
To this end, we set a regulated dc power supply (TAKASAGO, LTD.; KX-210L) for source 1 and a high-speed bipolar power amplifier of four-quadrant capability (NF Corporation; HSA4011) for source 2. 
The load consists of a resistor of $R_\mathrm{L} = 10\,\Omega$ and a capacitor of $C_\mathrm{L} = 9.9\,\mathrm{mF}$. 
The equivalent resistors, which represent the sum of the on-resistance across the routers other than the explicitly modeled ones, are set at $R_\mathrm{eq1} = R_\mathrm{eq2} = 3.3\,\Omega$. 
The time duration of one bit of a power packet is set at $4.0\,\mu\mathrm{s}$. 
The bit length of a payload is set at $240\,\mathrm{bit}$, leading to the duration of $960\,\mu\mathrm{s}$. 
Taking into account the 6-bit header and the 4-bit footer, the total bit length of a power packet is $250\,\mathrm{bit}$, and the corresponding time duration is $1.0\,\mathrm{ms}$. 
Based on these parameters, we derive the optimal design of $Q$ following the optimization procedure in \cite{Azuma.Sugie-2008a} as $A_Q = 0.9593$, $B_Q = 0.03060$, $C_Q = -31.34$. 

We set the reference input $\{u(k)\}$ as a triangular waveform
\begin{equation}
    u(k) = b + a \cdot \mathrm{tri}(2\pi k/K)
\end{equation}
where $a = 4.0\,\mathrm{V}$, $b = 8.0\,\mathrm{V}$, $K = 125$, and $\mathrm{tri}(\theta)$ is a triangular wave function of period $2\pi$ that takes 0 when $\theta = 2n\pi$ ($n=0,1,2,\dots$) and 1 when $\theta = m\pi/2$ ($m=1,3,5,\dots$). 
We adopt the reference waveform that incorporates both increasing and decreasing trends, which creates bidirectional power flow by charging and discharging stored energy of the load. 

\subsection{Results}

\begin{figure}[tb]
    \centering
    \includegraphics[width=9cm]{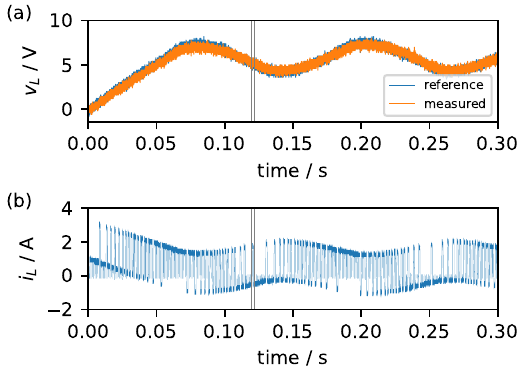}
    \caption{Experimental result of load voltage regulation. (a) Load voltage and its reference. (b) Load current (positive value for inflow direction).}
    \label{fig:overview}
\end{figure}

\fref{fig:overview} shows the result of load voltage regulation. 
\fref{fig:overview}~(a) depicts the load voltage with and without quantization. 
The load voltage without quantization was calculated numerically for reference with the model and parameters set above in the load-side router's controller and output through the controller's analog output port. 
The good agreement of the two lines indicates that the packetized power supply achieved successful regulation of load voltage in the sense of comparison with the continuous counterpart for reference. 

\fref{fig:overview}~(b) depicts the measured current, where the direction of inflow to the load is defined as positive. 
The bidirectional power flow is confirmed by observing the current of both positive and negative values.
In particular, by observing the current waveform together with changes in load voltage, it can be seen that the direction of power transmission is determined by changes in the stored energy of the load.
When the load voltage increases, the higher voltage source is selected more often, resulting in positive power transmission on average. 
When the load voltage decreases, the power source of lower voltage (battery) is selected more often, resulting in the negative (in regenerative direction) power transmission on avarage.

\begin{figure}[tb]
    \centering
    \includegraphics[width=9cm]{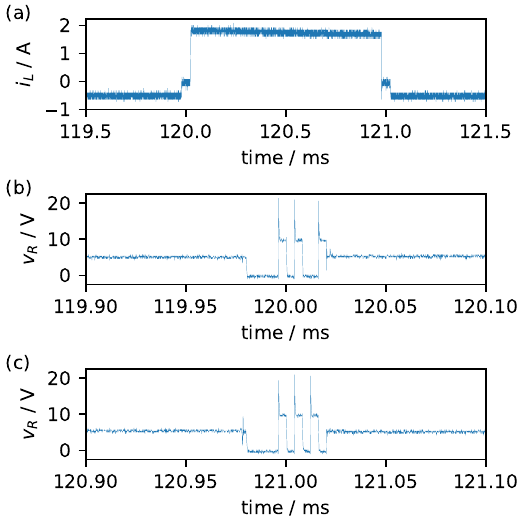}
    \caption{Enlarged view of the experimental results. (a) Load current in time span between gray vertical lines in Fig.~\ref{fig:overview}. (b) Tag waveform around $t=120\,\mathrm{ms}$. (c) Tag waveform around $t=121\,\mathrm{ms}$. }
    \label{fig:tag}
\end{figure}

Then, we observe the details of the signal transfer by information tags and the resulting power source selection. 
Since the time span presentation in Fig.~\ref{fig:overview} is not suitable to observe the details per packet, we refer to the enlarged view of the results presented in Fig.~\ref{fig:tag}. 
Fig.~\ref{fig:tag}~(a) presents the the enlarged view of the load current in $t\in[119.5\,\mathrm{ms},121.5\,\mathrm{ms}]$, where the header tags of two consecutive packets are depicted. 
The sign of the current indicates that the power packet beginning at around $t=120\,\mathrm{ms}$ is supplied by source 1, and that the power packet beginning at around $t=121\,\mathrm{ms}$ is regenerated to source 2. 
Indeed, Fig.~\ref{fig:tag}~(b)~and~(c) indicate that the headers of the former and latter power packets are \texttt{101001} and \texttt{101010}, respectively. 
These results ensure that the router selected the designated power source according to the information tag. 

\section{Discussions}
\label{sect:discuss}

In this paper, we proposed the method for the upstream allocation of the packetized bidirectional power demand of a dynamical load. 
The proposed method realized the seamless connection in the flow of both information and power and integrated them as power packets. 
We demonstrated the operation through experiments, presenting the successful allocation of load demand to both the pure source and the battery of source and sink capability. 

Although we focused on a minimal setup of two sources and one load in this paper, the proposed scheme can be applied to systems of more elements. 
First, the load demand allocation can easily be extended to three or more sources as follows. 
The software modification required is to expand the number of possible voltage levels in (\ref{eq:vlevel}).
The hardware requires to increase the number of ports in the source-side router.
Second, the proposed method can also be applied to multiple loads in the network. 
As illustrated in Section~\ref{sect:method}, the proposed scheme operates on a load-by-load basis, so simply applying the same algorithm to each load works.
Lastly, the load can be of any type if its numerical model is available. 
We considered a capacitive load in the setup to represent the capability of energy storing and discharging, and it can be replaced by e.g. an inductive load such as an electric motor. 

In the remainder of this section, we provide some discussions on the comparison of the proposed method with related works. 
The comparison highlights the standpoint and novelty of the present paper among related works. 

The concept of power packetization was first proposed in \cite{Toyoda.Saitoh-1998} in the 1990s and has been studied by independent groups around the world to date, e.g. \cite{He.etal-2008,Eaves-2012,Stalling.etal-2012,Gelenbe.Ceran-2016,Sugiyama.etal-2017,Almassalkhi.etal-2022}. 
Most of them involve information and power processing in separate layers and employ virtual tagging.
However, in situations where there are not enough adjustable power sources, the discrepancy of the information from physics at each instance can lead to serious accidents.
This is the reason why we adopt the packetization based on physical tagging. 

The coincidence of cyber and physical quantities distinguishes the proposed packet-based system from other converter-based approaches. 
In the context of bidirectional power flow management induced by batteries and renewable sources integration, the multiconverter architecture \cite{Emadi.Ehsani-2001} or the use of a multiport power converter \cite{Bhattacharjee.etal-2019} have been the main approaches.
These architectures, although there are variations for circuit topologies, in essence mix all the incoming or outgoing power at the DC and/or AC link. 
As mentioned in the Introduction, if there is a large buffer and/or source, they can serve as an absorber of the fluctuating gap between supply and demand.
Although they are often available in grid-tied applications, their introduction is costly and sometimes hindered by weight or size restrictions in the target applications of this paper. 
The proposed packet-based approach tackles this issue by completely dividing the delivery of each power packet. 
The effect of the fluctuating output of renewables stays only within the specific packet and is separated from others. 
Furthermore, the load packetization algorithm can compensate for the effect of fluctuation by changing the subsequent packet sequence because the first equation in (\ref{eq:quantizer}) contains the memory of the mismatch between the ideal and actual power supply. 

For the hardware setups of this paper, we have exploited the previously proposed method of circuit implementation. 
The original (unidirectional) power packet router was introduced in \cite{Takahashi.etal-2015}, and then was extended to accommodate bidirectional power packet flow in \cite{Yoshida.etal-2020}. 
The original router included the signal reading module, and the signal generation module was first introduced in \cite{Katayama.Hikihara-2020a}. 
The bidirectional router proposed in \cite{Yoshida.etal-2020} employed bidirectional switches, which enabled the bidirectional current flow, although the work did not discuss bidirectional load demand fulfillment by a packetized form of information and power transfer. 
Based on these hardware setups, the present work proposes a unified system solution for the upstream allocation of a bidirectional load demand. 

The concept of upstream load flow was introduced in a traditional power system analysis as a means to trace the load flow in terms of the contribution of the sources\cite{Bialek-1996}. 
Active allocation of upstream demand in a packetized form was first discussed in \cite{Nawata.etal-2016}. 
The work was limited to a static (pure resistive) and thus unidirectional load with power demand expressed on a time-average basis. 
On the other hand, this paper proposes a method that can handle a dynamic load whose demand is expressed as an instantaneous value. 
Furthermore, the previous work \cite{Nawata.etal-2016} analyzed the system using simplified numerical simulations and omitted the details of the routing operation on the physical layer. 
The proposed method presents how to deal with power and information processing in the physical layer. 

Regarding the load-demand representation in a packetized form, the present study extends our previous proposal in \cite{Takahashi.etal-2016a, Mochiyama.Hikihara-2019a}.
First, as mentioned in the Introduction, the previous works did not integrate the communication for this representation with the power packet transfer. 
The present work eliminates the limitation and realizes the full integration by the routing method developed in Section~\ref{sect:routing}. 
Furthermore, the previous works focused on a case where one source supplies one load, and thus the load regulation therein was recognized as a chopping operation of one power source with only two possible states of the circuit: whether the source is connected to the load or not. 
This setup implicitly limited the flow of current to unidirectional. 
The framework presented in Section~\ref{sect:src_select} overcomes the limitations above by extending the selection of power sources to different voltage levels, one for source operation and the other for sink operation. 


\bibliographystyle{IEEEtran}
\bibliography{hiwatashi}

\end{document}